\documentclass[english,twocolumn]{revtex4}
\usepackage[T1]{fontenc}
\usepackage[latin9]{inputenc}
\usepackage{amsmath}
\usepackage{graphicx}

\makeatletter
\@ifundefined{textcolor}{}
{%
 \definecolor{BLACK}{gray}{0}
 \definecolor{WHITE}{gray}{1}
 \definecolor{RED}{rgb}{1,0,0}
 \definecolor{GREEN}{rgb}{0,1,0}
 \definecolor{BLUE}{rgb}{0,0,1}
 \definecolor{CYAN}{cmyk}{1,0,0,0}
 \definecolor{MAGENTA}{cmyk}{0,1,0,0}
 \definecolor{YELLOW}{cmyk}{0,0,1,0}
}

\usepackage{lmodern}
\usepackage[T1]{fontenc}

\makeatother

\usepackage{babel}
\begin{document}

\title{Mode spectrum and temporal soliton formation \\in optical microresonators}

\author{T. Herr$^{1}$, V. Brasch$^{1}$, J.D. Jost$^{1}$, I. Mirgorodskiy$^{1,2}$,
G. Lihachev$^{2}$, M.L. Gorodetsky$^{2,3}$, T.J. Kippenberg$^{1}$}

\email{tobias.kippenberg@epfl.ch}

\selectlanguage{english}%

\affiliation{$^{1}$Ecole Polytechnique Fédérale de Lausanne (EPFL), 1015, Lausanne,
Switzerland}

\affiliation{$^{2}$Faculty of Physics, M.V.Lomonosov Moscow State University,
Moscow, 119991, Russia}

\affiliation{$^{3}$Russian Quantum Center, Skolkovo, 143025, Russia}
\begin{abstract}
The formation of temporal dissipative solitons in optical microresonators
enables compact, high repetition rate sources of ultra-short pulses
as well as low noise, broadband optical frequency combs with smooth
spectral envelopes. Here we study the influence of the resonator mode
spectrum on temporal soliton formation. Using frequency comb assisted
diode laser spectroscopy, the measured mode structure of crystalline
MgF$_{2}$ resonators are correlated with temporal soliton formation.
While an overal general anomalous dispersion is required, it is found
that higher order dispersion can be tolerated as long as it does not
dominate the resonator's mode structure. Mode coupling induced avoided
crossings in the resonator mode spectrum are found to prevent soliton
formation, when affecting resonator modes close to the pump laser.
The experimental observations are in excellent agreement with numerical
simulations based on the nonlinear coupled mode equations, which reveal
the rich interplay of mode crossings and soliton formation.
\end{abstract}
\maketitle
Temporal dissipative solitons \cite{Lugiato1987,Wabnitz1993,Leo2010}
can be formed in a Kerr-nonlinear optical microresonator \cite{Herr2012b}
with anomalous dispersion that is driven by a monochromatic continuous
wave pump laser. These temporal solitons are sech$^{2}$-shaped ultra-short
pulses of light circulating inside the microresonator, where the temporal
width of the solitons is fully determined by the resonator dispersion
and nonlinearity as well as the pump power and pump laser detuning
\cite{Herr2012b,Coen2013b}. It has been shown that the pump laser
parameters can be used to control the number of solitons circulating
in the microresonator. In particular the single soliton state, where
one single soliton is circulating continuously inside the resonator,
is of high interest for applications. In the time domain soliton formation
in microresonators allows for the generation of periodic ultra-short
femto-second pulses, which in the frequency domain correspond to a
frequency comb spectrum with smooth sech$^{2}$-shaped spectral envelope.
The free spectral range (FSR) of the resonator, typically in the range
of tens to hundreds of GHz, determines the pulse repetition rate (equivalent
to the frequency comb line spacing). Soliton formation is related
to four-wave mixing based frequency comb generation in microresonators
\cite{DelHaye2007,Savchenkov2008c,Grudinin2009,Razzari2010,Levy2010,Foster2011a,Papp2011,Kippenberg2011,Liang2011,Wang2013},
where low and high noise operating regimes \cite{Ferdous2011,Papp2011,Herr2012} have
been identified. Here, techniques such as $\delta-\Delta$-matching
\cite{Herr2012}, self-injection locking \cite{Li2012,DelHaye2013}
or parametric seeding \cite{Papp2013} can be used to achieve low
noise operation. In contrast to these low noise four-wave mixing based
combs (also termed Kerr combs), the transition to the soliton regime
\cite{Herr2012} offers a unique combination of features, such as
intrinsic low noise performance, direct pulse generation in the microresonator
\cite{Peccianti2012,Herr2012b,Saha2012a}, and smooth spectral envelope
as shown in Figure 1. These properties are critical to applications
in e.g. telecommunications \cite{Levy2012,Wang2012,Pfeifle2013}, low phase
noise microwave generation \cite{Savchenkov2004,Li2012}, precision spectroscopy
as well as frequency metrology \cite{Udem2002,Cundiff2003}.
\begin{figure}[t]
\label{fig:solitonComb}
\includegraphics[scale=0.6]{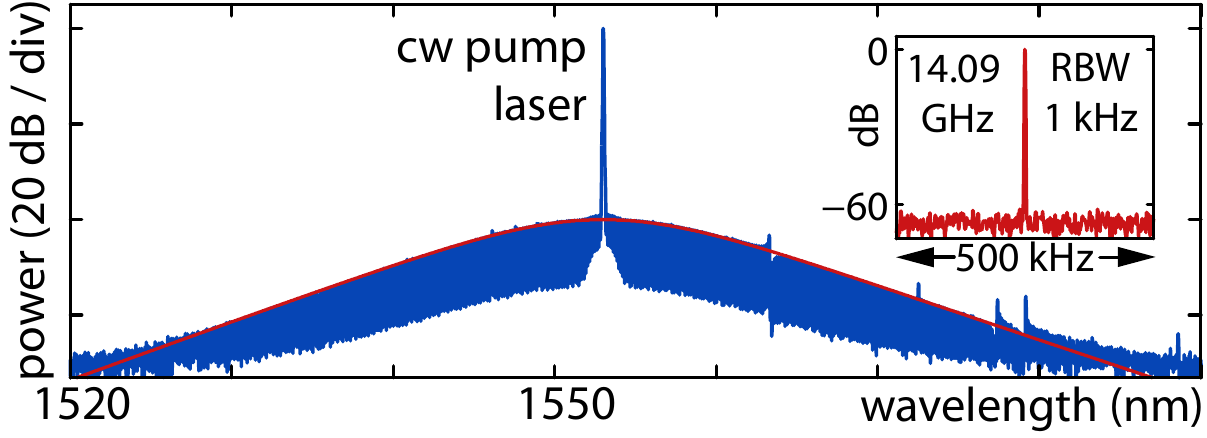}
\caption{Measured optical spectrum with smooth sech$^{2}$-shaped spectral
envelope (red line) of a single temporal soliton generated in a continuous
wave laser driven crystalline high-finesse MgF$_{2}$ microresonator.
The spectral 3-dB width of 13 nm (1.62 THz) corresponds to a soliton
pulse duration of 194 fs (full width at half maximum). The cw pump
power is $\sim3$0 mW at a wavelength of 1552 nm. The inset shows
the resolution bandwidth (RBW) limited RF signal at a frequency of
14.09 GHz corresponding to the soliton pulse repetition rate.}
\end{figure}
\begin{figure}[t]
\includegraphics[scale=0.65]{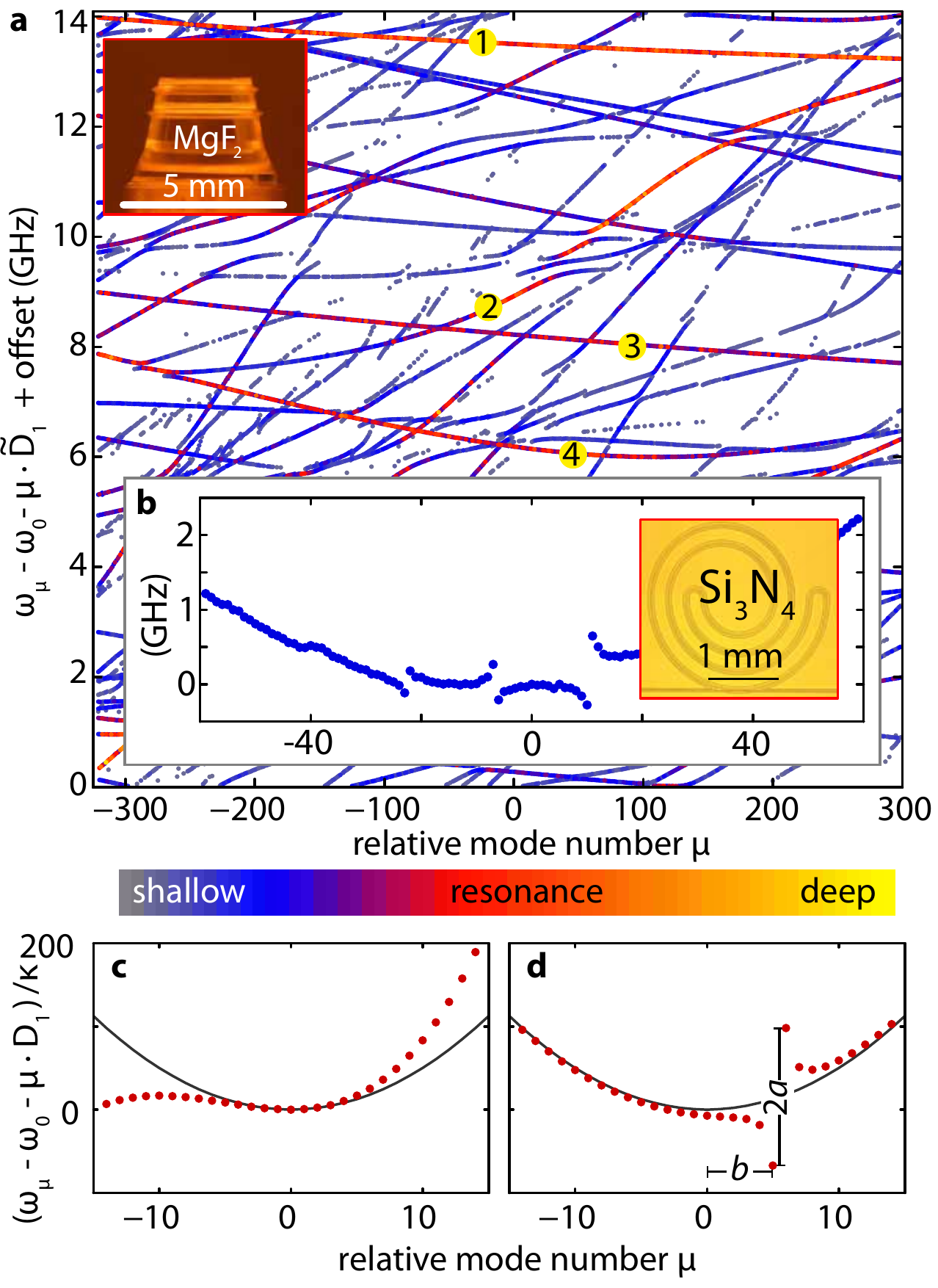}\label{fig:modeStructure}
\caption{(a) Mode structure of a MgF$_{2}$ resonator with linewidths in the
range of 50 to 500 kHz and an approximate FSR of 14.09 GHz, as measured
by frequency comb assisted diode laser spectroscopy\cite{DelHaye2009a}.
Dots forming a continuous line represent a particular mode family.
Different free spectral ranges correspond to different slopes of the
lines, whereas dispersion and variation of the free spectral range
show as curvature and bending of the lines. The dispersion can be
strongly affected by mode crossings. Four specific mode families have
been numbered by yellow labels. The color codes the measured resonance
depth and helps to track particular mode families. (b) Comparable
measurement of the fundamental TM11 mode in a Si$_{3}$N$_{4}$ microresonator
with a resonator linewidth of $350$ MHz and approximate FSR of 76
GHz (consisting of a 800 nm high and 2 $\mu$m wide Si$_{3}$N$_{4}$waveguide
embedded in SiO$_{2}$). The mode family shows signs of mode coupling
to other mode families. (c) Illustration of higher order dispersion
with $D_{3}>0$. The gray line indicates anomalous dispersion described
by $D_{2}$ only. (d) Illustration of mode coupling induced mode
frequency shift altering the dispersion properties locally. A simple
parametrization using magnitude $a$ and position $b$ of the avoided
mode crossing can be used for numerical modeling. }
\end{figure}
Temporal dissipative solitons in microresonators rely on the balance
between Kerr-nonlinearity and anomalous group-velocity dispersion
in the presence of a monochromatic pump laser and loss. Theory predicts
that soliton formation is possible in any Kerr-nonlinear microresonator
with anomalous dispersion for sufficiently high pump power \cite{Leo2010,Matsko2011a,Herr2012b,Coen2013,Chembo2013,Lamont2013}.
While high effective Kerr-nonlinearity and efficient non-linear frequency
conversion is routinely achieved in a wide variety of resonator materials
and geometries \cite{DelHaye2007,Savchenkov2008c,Grudinin2009,Razzari2010,Levy2010,Foster2011a,Papp2011,Kippenberg2011,Liang2011,Wang2013},
soliton formation has so far only been unambiguously demonstrated
in MgF$_{2}$ microresonators \cite{Herr2012b} where the characteristic
spectral sech$^{2}$-shape envelope has been observed (cf. Figure
1). In this work we show that the decisive requirement for the generation
of solitons is the anomalous resonator dispersion, which, in state
of the art resonators, is affected by higher order dispersion and
coupling of the optical modes \cite{Carmon2008,DelHaye2009a,Savchenkov2012,Grudinin2013}. Understanding the formation of temporal
dissipative solitons in the context of a complex mode-structure is
not only an interesting and open scientific question, but essential
to reliably reproduce soliton formation in other microresonator platforms
such as SiN \cite{Levy2010,Foster2011a}. 

The dispersion of a microresonator can be described in terms of its
resonance frequencies $\omega_{\mu}$ using the parameters $D_{1}$,
$D_{2}$, $D_{3}$ etc., which correspond to the free spectral range
(FSR) in radians, the second order dispersion and higher order dispersion
parameters, respectively \cite{Savchenkov2011,Herr2012}

\begin{equation}
\omega_{\mu}=\omega_{0}+D_{1}\mu+\tfrac{1}{2}D_{2}\mu^{2}+\tfrac{1}{6}D_{3}\mu^{3}+...\hspace{1em}.\label{eq:mode_expansion}
\end{equation}

Here, $\mu$ denotes the relative mode number with respect to the
pump (designated by $\mu=0$). The parameter $D_{2}$ is related to
the often employed group velocity dispersion (GVD) parameter $\beta_{2}$
via 
\begin{equation}
D_{2}=-\frac{c}{n}D_{1}^{2}\beta_{2}\hspace{1em}.
\end{equation}
A positive $D_{2}$ corresponds to an anomalous resonator dispersion
leading to a parabolic deviation of the resonance frequencies from
an equidistant $D_{1}$-spaced grid (cf. grey curve in Figure 2c,d).
This anomalous dispersion can be modified by higher order terms such
as $D_{3}$ as illustrated schematically in Figure 2c. The dispersion
coefficients $D_{2}$, $D_{3}$, etc. are typically estimated either
analytically \cite{Schiller1993,Gorodetsky1994,Gorodetsky2006,Gorodetsky2007,Demchenko2013}
or numerically \cite{Oxborrow2007,DelHaye2009a,Riemensberger2012,Saha2012b}
by taking material and the geometrical dispersive effect into account.
It is well known, that the coupling (e.g. via scattering) between
mode families can additionally modify the mode frequencies \cite{Carmon2008,DelHaye2009a,Savchenkov2012,Grudinin2013}
and lead to avoided crossings (illustrated schematically in Fig. 2d).
In a simplified model the effect of a mode-crossing can be parametrized
by its magnitude $a$ and position $b$ (details below). To investigate
the dispersion requirements for soliton formation in an experimental
system, broadband frequency comb assisted scanning laser spectroscopy
\cite{DelHaye2009a} is used to precisely characterize the complex
mode structure of a MgF$_{2}$ microresonator (FSR 14.09 GHz) \cite{Liang2010,Grudinin2012,Wang2013}
over a spectral span exceeding 8 THz (including more than 600 FSR
and several tens of mode families). From the recorded, frequency comb
calibrated transmission spectra the resonance frequencies are determined.
The measured mode structure is visualized in Figure 2 using a 2-dimensional
representation. Here for each detected mode family and relative mode
number $\mu$, the mode frequency is given with respect to a common
equidistant frequency grid with a spacing of $\tilde{D_{1}}/2\pi=14.095$
GHz (close to the approximate average FSR, but chosen arbitrarily).
The high number of optical modes and their complex mode-structure
allow for investigation of different regimes of resonator dispersion
and deriving empirical criteria for soliton formation. 
\begin{figure}
\includegraphics[scale=0.55]{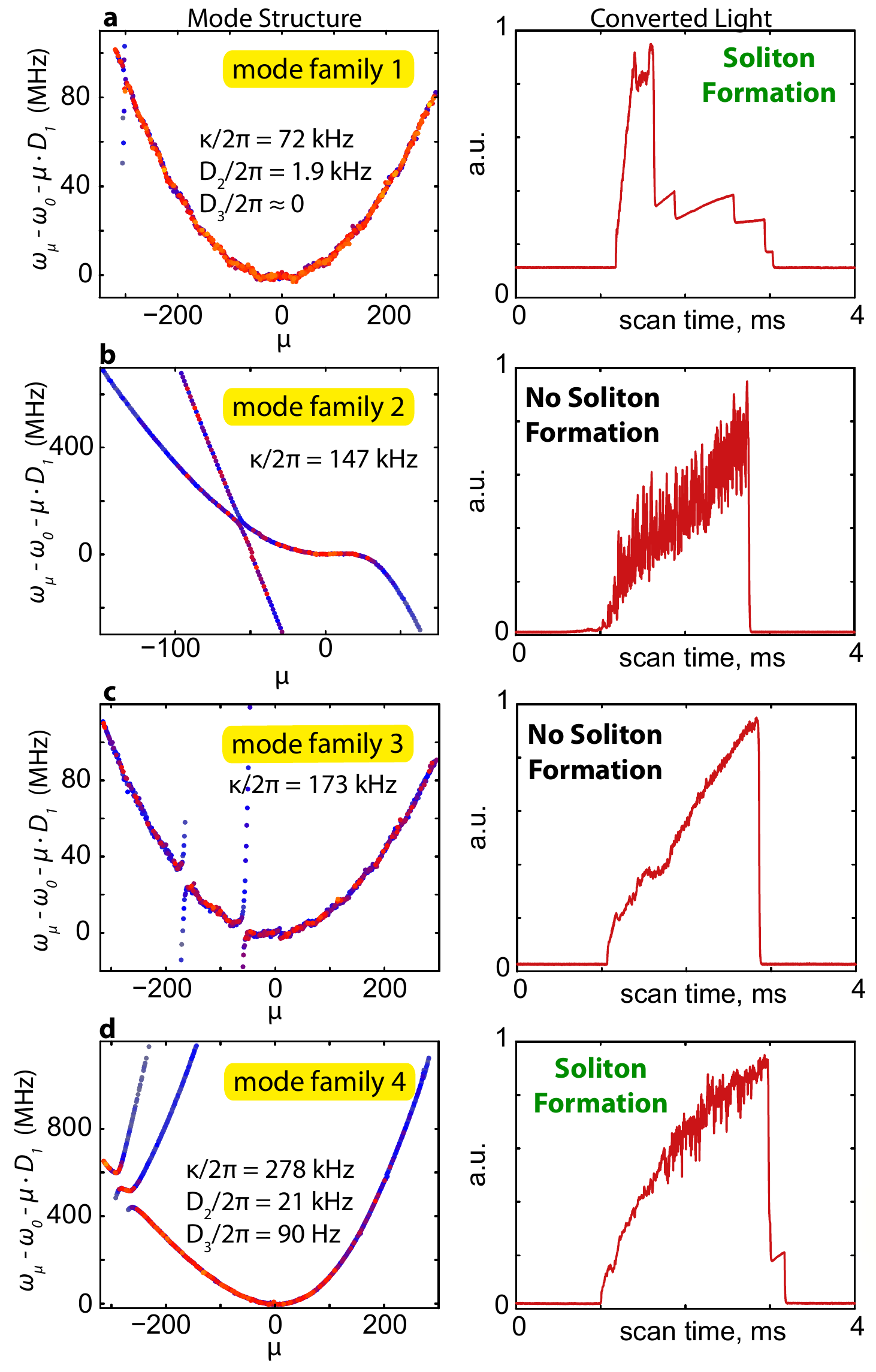}\label{fig:6ResSteps}\caption{Experimental investigation of soliton formation in different dispersion
scenarios in a MgF$_{2}$ microresonator. (a and d) Soliton formation
is observed in the resonance families 1 and 4 of Figure 1, which show
an almost ideal $D_{2}>0$ dominated anomalous dispersion. The soliton
formation is detected by observation of the step signature in the
converted light signal. (b and c) Soliton formation is not observed
in the mode families 2 and 3 where strong deviations from $D_{2}$
dominated dispersion are present. The coupled pump power is $\mathcal{O}(1\mathrm{mW)}$
at a wavelength of 1552~nm.}
\end{figure}
\begin{figure*}[!t]
\includegraphics[scale=0.7]{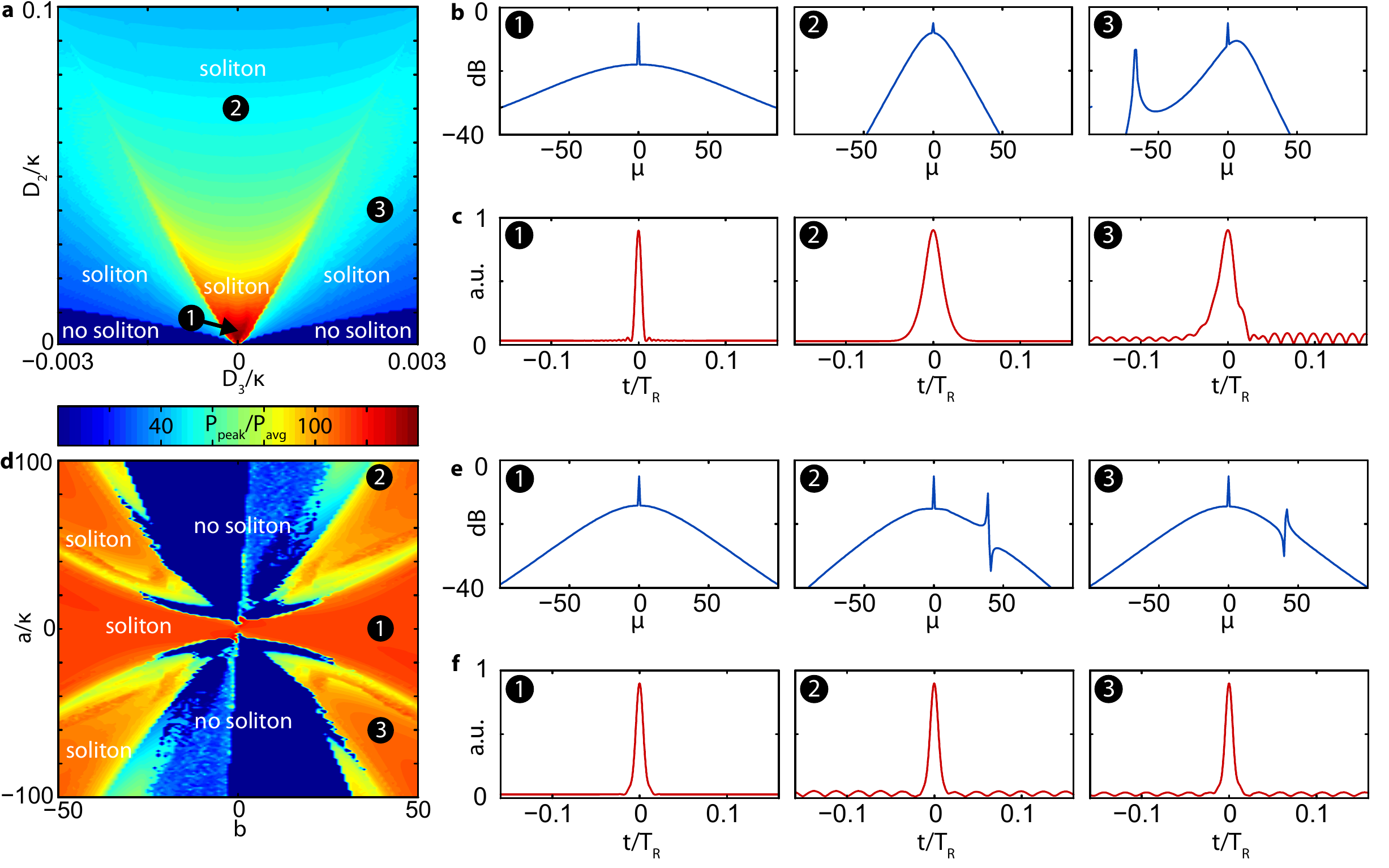}\label{fig:numRes}\caption{Numerical investigation of soliton formation in different dispersion
scenarios. (a) The ratio $P_{\mathrm{peak}}/P_{\mathrm{avg}}$ of
peak to average intracavity power is used as an indicator of soliton
formation (which results in high peak power) for different combinations
of $D_{2}$ and $D_{3}$. (b) Optical spectra for different parameters
simulated in panel a (1,2,3) show narrow spectral width for high values
of $D_{2}$ and asymmetric spectra, as well as dispersive wave phenomena
(peak at $\mu=-65$), for nonzero $D_{3}$. (c) Temporal field envelope
inside the microresonator corresponding to the spectra shown in panel
b. The soliton pulse duration lengthens for larger values of $D_{2}$.
Oscillatory features in the background field appear for nonzero $D_{3}$.
(d) The high ratio of peak to average power is used as an indicator
of soliton formation for different situations characterized by an
avoided mode crossings that is parametrized by magnitude $a$ and
distance $b$ from the pump laser (cf. Figure 2d). (e) Optical spectra
for different simulation parameters in panel d (1,2,3) show the characteristic
'up-down' feature induced by an avoided mode-crossing. (f) Temporal
field envelope inside the microresonator corresponding to panel e.
The pulses have the same duration but oscillatory features in the
background field appear in the presence of avoided mode crossings.}
\end{figure*}
Mode families with different free spectral range (corresponding to
different radial and meridional mode numbers) that cross other mode
families or that show modified dispersion due to avoided mode crossings
can be observed. Some mode families exhibit normal other mode families
anomalous dispersion. Moreover, mode coupling can locally alter the
dispersion characteristics of a mode. The inset in Figure 2 shows
that the effects of interacting modes are not only present in the
case of a crystalline MgF$_{2}$ microresonator but also occur in
a Si$_{3}$N$_{4}$ microresonator that due to its small cross-section
supports only few higher order modes.

Next, the dispersion of four individual mode families is related to
their potential of supporting temporal solitons. All four mode families
allow for efficient non-linear frequency conversion. The latter is
measured by detecting the parametrically frequency-converted laser
light (i.e. the out-coupled optical comb spectrum with the pump wavelength
filtered out, using a narrow-band fiber-Bragg grating) with a photo-detector.
The converted light signal also provides direct means of detecting
soliton formation. The latter exploits the fact that the formation
of solitons coincides with discrete steps in the converted light signal
(cf. Figure 3a,d, right column) that are observed when the pump laser
is scanned over the resonance (see supplementary information in ref.
\cite{Herr2012b}). The signals of the converted laser light for the
four pumped modes ($\mu=0$) are shown in Figure 3 (right column)
as function of the scan time. Despite significant nonlinear frequency
conversion (and associated formation of comb-like broadband spectra)
in all four mode families, only two mode families (1 and 4) exhibit
soliton formation. To understand the reason, the dispersion properties
of all four modes are investigated in detail based on the data shown
in Figure 2a. Figure 3 (left column) shows the deviation of the resonance
frequencies of the individual mode families from an equidistant frequency
grid defined by the FSR (i.e. $D_{1}/2\pi$) of the mode family at
the pumped resonance $\mu=0$. A perfectly anomalous dispersion, i.e.
$D_{2}>0$ and vanishing higher order dispersion terms correspond
to a convex, parabolic curve. This case is closely realized for mode
family 1 ($D_{2}/2\pi=1.9$ kHz, $D_{3}/2\pi\approx0$), which also
shows the characteristic step signature \cite{Herr2012b} of soliton
formation in Figure 3. Mode family 2 (Figure 3b) is not characterized
by an anomalous dispersion and does not show signs of soliton formation.
Mode family 3 (Figure 3c) exhibits an overall anomalous dispersion
that is however disturbed locally by two avoided mode-crossings in
the spectral proximity of the pumped mode. As a result no solitons
are formed. In contrast, solitons are generated in mode family 4 (Figure
3d), which in addition to a dominating anomalous dispersion ($D_{2}/2\pi=21$
kHz) is characterized by a noticeable higher order contribution to
its dispersion ($D_{3}/2\pi=90$ Hz). Moreover, the smooth dispersion
curve is disturbed by two avoided mode crossings well separated in
terms of mode number from the pumped mode. These measurements reveal
that the formation of soliton is robust against a certain contribution
of higher order dispersion as well as local mode frequency shift induced
by mode coupling. However, an overall dispersion that is not generally
anomalous (as in the case of mode family 2) or avoided mode-crossings
too close to the pumped mode (as in the case of mode family 3) does
prevent soliton formation. Note that a high optical finesse (i.e.
narrow linewidth) is not a decisive requirement for soliton formation.
Indeed the two soliton generating mode families (1 and 4) posses intermediate
linewidths of $\kappa/2\pi=72$ kHz and 173 kHz compared to the mode
families that do not support soliton formation (2 and 3) where $\kappa/2\pi=147$
kHz and 278 kHz. In short, the above experiments, correlating soliton
formation with broadband precision dispersion mesurements, suggest
that higher order dispersion and avoided mode-crossing are features
that can prevent soliton formation. In the following we utilize numerical
simulations to test and substantiate this experimentally motivated
hypothesis. 

The nonlinear physics in optical microresonators can be accurately
modeled in the time or frequency domain, or via split-step methods
combining both domains. Here we employ a split-step method implemented
as an extension to the non-linearly coupled modes approach\cite{Chembo2010,Chembo2013,Hansson2014}.
A key advantage of this method is that it allows for the definition
of arbitrary mode frequencies $\omega_{\mu}$, enabling complex mode-structures
to be readily simulated. To investigate whether a particular mode
structure $\omega_{\mu}$ ($\mu=0,\pm1,...$) allows for soliton formation,
pump laser scans that can lead to the formation of solitons \cite{Herr2012b}
are numerically simulated. To ensure deterministic computational evolution
into a single soliton state the unperturbed analytical single soliton
waveform \cite{Herr2012b} is used for seeding the simulation. For
each simulated pump laser scan the maximum ratio of peak to average
power inside the microresonator is computed. This ratio serves as
a reliable indicator of soliton formation inside the microresonator.
Throughout the simulation typical microresonator parameters of $\kappa/2\pi=1$
MHz, a FSR of $35$ GHz (roundtrip time $T_{R}\approx29$ ps), an
effective nonlinearity of $\gamma=4\times10^{-4}$m$^{-1}$W$^{-1}$
and a pump power of $100$ mW at 1.55 $\mu$m wavelength are assumed.
However all the simulations were performed in dimensionless units
\cite{Herr2012b} and may be rescaled to other required sets of parameters.

First, to study the effect of higher order dispersion the optical
modes are defined according to Equation \ref{eq:mode_expansion} with
varying $D_{2}$ and $D_{3}$ (Note that the offset $\omega_{0}$
and the linear term $D_{1}$ can be chosen arbitrarily). The numerical
results in Figure 4a,b,c show that nonzero values of $D_{3}$ require
a minimal magnitude of the coefficient $D_{2}$ to allow for soliton
formation (consistent with our experimental observations). The maximum
soliton peak intensities and shortest soliton pulse durations are
achieved for vanishing $D_{3}$ and small values of $D_{2}>0.$ Figure
4a is a contour plot of the peak resonator intensities as a function
of $D_{2}$ and $D_{3}$. It can be noted that the peak intensities
are invariant under change of sign in $D_{3}$. The higher order dispersion
$D_{3}$ causes the optical spectrum to become asymmetric and in the
simulation is seen to result in dispersive wave formation \cite{Erkintalo2012,Coen2013,Lamont2013}.

Second, the effect of avoided mode-crossings is studied in a simplified
model. Here the mode frequencies are defined according to 

\begin{equation}
\omega_{\mu}=\omega_{0}+D_{1}\mu+\tfrac{1}{2}D_{2}\mu^{2}+\frac{a/2}{\mu-b-0.5}\label{eq:mode_shifted}
\end{equation}
to phenomenologically mimic the effect of resonance frequency shifts
induced by an avoided mode crossing. The parameters $a$ and $b$
specify the magnitude of this frequency shift and distance of the
frequency from the pumped mode (cf. Figure 2d). Subtracting the value
$0.5$ in the denominator of Equation \ref{eq:mode_shifted} avoids
an infinite mode shift for $\mu=b.$ As a result the maximum resonance
frequency shift of $a$ occurs for the modes with modes numbers $\mu=b$
and $\mu=b+1$ towards smaller and higher frequency, respectively
(The validity of the simple model has been verified in much slower
direct simulation of mode coupling between two separate mode families,
which led to similar results). The results of the simulations for
various values of $a$ and $b$ are shown in Figure 4d,e,f. Here,
Figure 4d shows the peak power as a function of the strength and location
of the mode crossing. The contour plot exhibits a point symmetry,
reflecting the equivalence of the mode shifts defined by $\{+a,+b\}$
and $\{-a,-b\}$, respectively. While Figure 4d reveals a rich and
complex structure, it shows that generally, the larger the spectral
separation $b$ of the mode crossing from the pumped mode ($\mu=0$),
the higher the magnitude $a$ of the mode crossing can be without
preventing soliton formation. The presence of crossing manifests itself
in the optical spectrum as characteristic features,where the spectral
intensities are increased on one side of the avoided crossing and
decreased on the other cf. Figure 4e, trace 2 and 3 \cite{Grudinin2013}.
These features are evidenced experimentally in Figure 1c. Increasing
the magnitude of the mode-crossing $a$ to larger values eventually
inhibits the formation of solitons, in agreement with the experimental
observations.

In summary, we have shown experimentally and numerically that a $D_{2}$
dominated anomalous resonator dispersion as well as a low number of
mode crossings are essential prerequisites to the generation of temporal
dissipative solitons in microresonators. A low number of avoided mode
crossings can be achieved by reducing mode coupling and by designing
single mode resonators.

\textbf{Acknowledgements:} This work was supported by the
European Space Agency (V.B), a Marie Curie IIF (J.D.J), the Swiss
National Science Foundation (T.H.), the Eurostars program and the
DARPA QuASAR program, RFBR grant 13-02-00271 and state contract 07.514.12.4032
(M.L.G). The research leading to these results has received funding
from the European Union Seventh Framework Programme (FP7/2007-2013)
under grant agreement no. 263500.

\bibliographystyle{apsrev}
\bibliography{library}

\end{document}